\documentclass[]{mn2e}

\usepackage{graphics}
\usepackage{rotating}
\usepackage{amsfonts}

\def\apj{ApJ}
\def\aap{A\&A}

\title{Gravitational wave radiation from the coalescence of white dwarfs}

\author[P. Lor\'en--Aguilar et al.]{P. Lor\'en--Aguilar$^{1,2}$,
                                    J. Guerrero$^{1,2}$
                                    J. Isern$^{1,2}$ 
                                    J.A. Lobo$^{1,3}$ and 
                                    E. Garc\'\i a--Berro$^{1,4}$\\
$^1$Institut  d'Estudis Espacials  de Catalunya,  Edifici  Nexus, Gran
    Capit\`a 2-4, 08034 Barcelona, Spain \\
$^2$Institut de Ci\`encies de l'Espai, C.S.I.C\\
$^3$Departament  de F\'\i sica  Fonamental, Universitat  de Barcelona,
    c/Mart{\'\i} i Franqu\`es 1, 08028 Barcelona, Spain.\\
$^4$Departament  de F\'\i sica Aplicada,  Universitat  Polit\`ecnica  de
    Catalunya,  Escola Polit\`ecnica Superior de Castelldefels, \\ Avda. 
    del Canal Ol\'\i mpic s/n, 08860 Castelldefels, Spain.}

\begin{document}

\maketitle

\begin{abstract}
We compute the emission of gravitational radiation from the merging of
a close  white dwarf binary system. This  is done for a  wide range of
masses  and compositions  of the  white dwarfs,  ranging  from mergers
involving two  He white dwarfs, mergers  in which two  CO white dwarfs
coalesce to  mergers in which a  massive ONe white  dwarf is involved.
In doing so we follow the  evolution of binary system using a Smoothed
Particle Hydrodynamics  code.  Even though the  coalescence process of
the white  dwarfs involves  considerable masses, moving  at relatively
high  velocities with  a high  degree of  asymmetry we  find  that the
signature  of  the merger  is  not very  strong.   In  fact, the  most
prominent feature  of the  coalescence is that  in a  relatively small
time  scale (of  the order  of the  period of  the last  stable orbit,
typically  a  few minutes)  the  sources  stop emitting  gravitational
waves.  We also discuss  the possible implications of our calculations
for the  detection of the  coalescence within the framework  of future
space-borne interferometers like LISA.
\end{abstract}


\section{Introduction}

Gravitational waves are a direct  consequence of the General Theory of
Relativity.  Many  efforts have been done  so far to  detect them, but
due to the intrinsic experimental difficulties involved in the process
of detection and in the data  analysis no definite result has been yet
obtained.  Supernova  core collapse, binary  systems involving compact
objects, like double black holes, double neutron stars or double white
dwarfs are,  amongst others, promising sources  of gravitational waves
--- see  Schutz (1999)  for  a comprehensive  review  on the  subject.
Moreover,  with the advent  of the  current generation  of terrestrial
gravitational  wave detectors,  like LIGO  (Abramovici et  al.  1992),
VIRGO (Acernese et  al.  2004), GEO600 (Willke et  al.  2004), or TAMA
(Takahashi  et al.   2004),  and of  space-borne interferometers  like
LISA\footnote{\tt  http://lisa.jpl.nasa.gov}  (Bender  et  al.   1998,
2000), gravitational  wave astronomy will probably be  soon a tangible
reality.
 
As  already   mentioned,  one  of   the  most  promising   sources  of
gravitational waves  are galactic  binary systems containing  at least
one  compact  object.   Galactic   binaries,  such  as  neutron  stars
binaries,  cataclysmic binaries  or  close white  dwarf binaries,  are
guaranteed  sources for LISA  (Mironowsky 1965;  Evans, Iben  \& Smarr
1987), provided that the  sources are at sufficiently close distances.
In fact,  emission from galactic  close white dwarf binary  systems is
expected to  be the dominant  contribution to the background  noise in
the low frequency region, which ranges from $\sim 10^{-3}$ up to $\sim
10^{-2}$~Hz  (Bender et  al.  1998).   Additionally, from  very simple
considerations about the initial mass  funtion, it is easy to see that
galactic close white dwarf binaries must be quite common (Hils, Bender
\&  Webbink   1990)  and,  consequently,  if  the   amplitude  of  the
gravitational waves  is large enough  we should be able  to eventually
detect them  during the operation  of LISA.  Moreover, the  merging of
two white  dwarfs by emission  of gravitational radiation will  be the
final  destiny of  a good  fraction of  this type  of  binary systems.
Since during  the merging process  a sizeable amount  of gravitational
waves  is  expected  to be  produced  (Guerrero  et  al. 2004)  it  is
important  to  characterize  which  would be  the  gravitational  wave
emission of such  process and to assess the  feasibility of dectecting
them.

The process  of formation of  close white dwarf binaries  involves two
mass  transfer episodes  of  the  progenitor stars  when  each of  the
components  of  the  binary  system  evolves off  the  main  sequence.
Depending on when  during the lives of the  binary components the mass
transfer  episodes  occur  the  components  may  have  different  core
compositions.  In particular,  we may have He-He systems  with a total
mass $M_{\rm tot}  \la 0.75\, M_{\sun}$, He-CO for  those systems with
masss  within the  range $0.75\,  M_{\sun}  \la M_{\rm  tot} \la  1.45
M_{\sun}$,  CO-CO for  masses  larger than  $M_{\rm  tot} \sim  1.45\,
M_{\sun}$  and even He-ONe  or CO-ONe  systems when  one of  the white
dwarfs  is a  massive one.   Although the  astrophysical  scenarios in
which a merger of two white  dwarfs in a close binary system can occur
and their  relative frequencies have been relatively  well studied ---
see,  for instance,  Yungelson et  al.   (1994), and  Nelemans et  al.
(2001a,  2001b), and  references therein  --- the  process  of merging
itself has received little attention until very recently.  Indeed, one
of the  probable reasons  for this lack  of theoretical models  is the
heavy  computational   demand  involved   in  the  simulation   of  an
intrinsically   three-dimensional  phenomenon.    However,   in  sharp
contrast, the  coalescence of two  neutron stars has  been extensively
studied  --- see,  for instance,  Rosswog et  al.  (2000),  Rosswog \&
Davies (2002),  and Rosswog  \& Liebend\"orfer (2003),  and references
therein, for some of the most recent works on this subject.

In a  recent paper (Guerrero et  al.  2004) we  thoroughly studied the
merging of white  dwarf binary systems for a wide  range of masses and
compositions.   In doing  this  we used  an  up-to-date {\sl  Smoothed
Particle Hydrodynamics}  code. This method was first  proposed by Lucy
(1977) and, independently,  by Gingold  \& Monaghan (1977).   The fact
that  the method is  totally Lagrangian  and does  not require  a grid
makes   it   specially   suitable   for  studying   an   intrinsically
three-dimensional problem like the coalescence of two white dwarfs. In
the present paper we will not discuss the simulations presented there.
The interested  reader can find them  in mpeg format  at the following
URL: {\tt ftp://ftp.ieec.fcr.es/pub/astrofisica/SPH}. Instead, in this
paper we  compute the gravitational  wave signature expected  from the
merging of two  white dwarfs as obtained from  those SPH calculations.
The paper  is organized as follows.   In \S 2 we  briefly describe the
main ingredients  of our SPH  code, whereas in  \S 3 we  summarize the
physics of  the emission of gravitational  waves.  In \S  4 we discuss
the calibration and consistency checks we have done in order to assess
the reliability  of our results, whereas in  \S 5 and \S  6 we present
our results.  Finally,  in \S 7 we summarize our  results and draw our
conclusions.

\section{The Smoothed Particle Hydrodynamics code}

We  follow the  coalescence of  the binary  system using  a Lagrangian
particle numerical  code and,  more specifically, a  Smoothed Particle
Hydrodynamics (SPH) code. Our SPH code has been described at length in
Guerrero  et al.   (2004). However,  for the  sake of  completeness we
provide  here  a short  description  of  the  main input  physics  and
numerical techniques.   Our code follows closely  the prescriptions of
Benz  (1990),  where  the  basic  numerical  scheme  for  solving  the
hydrodynamic  equations can be  found. We  use the  standard polynomic
kernel of Monaghan \&  Lattanzio (1985).  The gravitational forces are
evaluated  using  an octree  (Barnes  \&  Hut  1986).  The  artificial
viscosity adopted in  this work is that of  Balsara (1995).  Regarding
the integration  method we use a  predictor-corrector numerical scheme
with variable time step (Serna, Alimi \& Chieze 1996), which turns out
to be  quite accurate.  The  adopted equation of  state is the  sum of
three components.   The ions  are treated as  an ideal gas  but taking
into account  the coulombian  corrections.  We have  also incorporated
the pressure of photons.   Finally, the most important contribution is
the pressure of degenerate electrons  which is treated as the standard
zero  temperature expression  plus the  temperature  corrections.  The
nuclear  network  adopted  here   (Benz,  Hill,  \&  Thielemann  1989)
incorporates 14 nuclei: He,  C, O, Ne, Mg, Si, S, Ar,  Ca, Ti, Cr, Fe,
Ni  and  Zn.   The  reactions  considered  are  captures  of  $\alpha$
particles,  and the  associated back  reactions, the  fusion of  two C
nuclei, and  the reaction between C  and O nuclei.  All  the rates are
taken from Rauscher \& Thielemann (2000). The thermal evolution of the
system is followed in two ways.   On the one hand the variation of the
internal energy is followed according to:

\begin{equation}
\frac{du_i}{dt}=\frac{P_i}{\rho_i^2}\sum_{j=1}^N m_j \vec{v_{ij}}
\vec{\nabla_i}W(r_{ij},h)
\end{equation}

\noindent where $W(r_{ij},h)$ is the  smoothing kernel and the rest of
the symbols have their usual  meaning.  The smooting length adopted in
in this work is the linear  average of the particles considered in the
corresponding calculation; that  is $h_{ij}=(h_i+h_j)/2$.On the other,
we simultaneously compute the temperature variation according to:

\begin{equation}
\frac{dT_i}{dt}=-\sum_{j=1}^N \frac{m_j}{(C_{\rm v})_j}
\frac{T_j}{\rho_i \rho_j}\Big[\big(\frac{\partial P}
{\partial t}\big)_\rho\Big]_j \vec{v_{ij}}\vec{\nabla_i}W(r_{ij},h)
\end{equation}

\noindent  If in  the  region  under study  the  temperature is  below
$6\times   10^8$~K   or  the   density   is   smaller  than   $6\times
10^3$~g/cm$^3$ we use equation (2)  otherwise we use equation (1).  We
have found that in this way energy is better conserved.
 
\section{Emission of Gravitational Waves}

We  compute  the  gravitational  wave  emission  in  the  slow-motion,
weak-field  quadrupole  approximation  (Misner  et  al.   1973).   The
dimensionless wave  strain, $h$, in the  transverse-traceless gauge is
given by:

\begin{equation} 
h^{\rm TT}_{jk}(t,\textbf{x})=\frac{2G}{c^4d}\frac{\partial ^2 
Q^{\rm TT}_{jk}(t-R)}{\partial t^2}
\label{sol_retarded}
\end{equation}

\noindent where $t-R=t-d/c$ is the  retarded time, $d$ is the distance
to the  observer and $Q^{\rm TT}_{jk}(t-R)$ is  the reduced quadrupole
moment of the mass distribution, which is given by:

\begin{equation} 
Q^{\rm TT}_{jk}(t-R)=\int{\rho(\textbf{x},t-R)
(x^jx^k-\frac{1}{3}x^2\delta_{jk})d^3x}
\label{mom_quadrupolar}
\end{equation}

\noindent The  rest of  the symbols have  their usual meaning.   It is
useful to express the time  derivative of the quadrupole moment in the
following way (Nakamura \& Oohara 1989):

\begin{equation} 
\ddot Q^{\rm TT}_{jk}(t-R)=P_{ijkl}(\textbf{N})\int{d^3 x\rho}
\big\lbrack 2v^kv^l-x^k\partial^l\phi -x^l\partial^k\phi\big\rbrack
\label{mom_derivative}
\end{equation}

\noindent where

\begin{eqnarray} 
P_{ijkl}(\textbf{N})& \equiv & (\delta_{ij}-N_iN_k)(\delta_{jl}-N_jN_l)\cr
&-&\frac{1}{2}(\delta_{ij}-N_iN_j)(\delta_{kl}-N_kN_l)
\label{pol_tensor}
\end{eqnarray}

\noindent  is the  transverse-traceless projection  operator  onto the
plane  ortogonal to  the  outgoing wave  direction, $\textbf{N}$,  and
$\phi$  is   the  gravitational  potential.   Now,   one  can  express
Eq.~(\ref{sol_retarded}) in the following way:

\begin{equation}  
h^{\rm TT}_{jk}(t,\textbf{x})=\frac{G}{c^4 d}\big(A_{+}(t,
\textbf{x})\textbf{e}_{+\, jk}+A_{\times}(t,\textbf{x})
\textbf{e}_{\times\, jk}\big)
\label{pol_decomposition}
\end{equation}

\noindent  where the polarization tensor coordinate matrices are defined
as: 

\begin{eqnarray} 
\textbf{e}_{+\, jk}&=&\frac{1}{\sqrt{2}}[(\textbf{e}_{x})_j
(\textbf{e}_{x})_k-(\textbf{e}_{y})_j(\textbf{e}_{y})_k]\cr
&&\\
\textbf{e}_{\times\, jk}&=&\frac{1}{\sqrt{2}}[(\textbf{e}_{x})_j
(\textbf{e}_{y})_k+(\textbf{e}_{y})_j(\textbf{e}_{x})_k],\nonumber
\end{eqnarray}

\noindent  the dimensionless amplitudes  $h_{+} \equiv  A_{+}/{d}$ and
$h_{\times} \equiv  A_{\times}/{d}$ are  the two independent  modes of
polarization in the transverse-traceless gauge, and the amplitudes are
respectively given by

\begin{equation}  
A_{+}(t,\textbf{x})= \ddot Q_{xx} - \ddot Q_{yy}, \;\;\; 
A_{\times}(t,\textbf{x})=+2\ddot Q_{xy}
\label{ATT1}
\end{equation}

\noindent for $i=0$, and 

\begin{equation}  
A_{+}(t,\textbf{x})= \ddot Q_{zz} - \ddot Q_{yy}, \;\;\;  
A_{\times}(t,\textbf{x})=-2\ddot Q_{yz}
\label{ATT2}
\end{equation}

\noindent  for $i=\pi/2$.  

In  our case we  have a  collection of  $n$ individual  SPH particles.
Consequently, Eq.~(\ref{mom_derivative}) must be discretized and it is
computed according to the following expression:

\begin{eqnarray} 
\ddot Q^{\rm TT}_{jk}(t-R)&\approx&  P_{ijkl}(\textbf{N})\sum^{n}_{p=1} 
m(p) \lbrack  2\textbf{v}^k(p)\textbf{v}^l(p)\cr
&+&\textbf{x}^k(p)\textbf{a}^l(p)+\textbf{x}^l(p)\textbf{a}^k(p) \rbrack
\label{mom_discretized}
\end{eqnarray}

\noindent  Where  $m(p)$  is  the  mass  of  each  SPH  particle,  and
$\textbf{x}(p)$,    $\textbf{v}(p)$     and    $\textbf{a}(p)$    are,
respectively, its position, velocity and acceleration.

\begin{figure}
\centering
\includegraphics[clip,width=250pt]{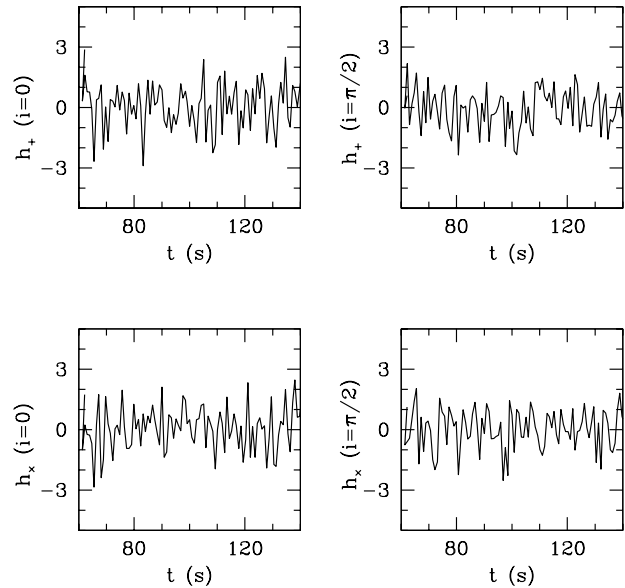}
\caption{Gravitational wave emission from an isolated, spherical star.
        The  dimensionless   strains  $h_{+}$  and   $h_{\times}$  are
        measured in  units of $10^{-25}$.  The source is located  at a
        distance of 10~kpc.}
\label{graf_zero}
\end{figure}

\section{Calibration and consistency checks}

We have  done two  tests, gravitational wave  emission from  a single,
isolated  star, and  gravitational wave  emission from  a  close white
dwarf binary  system in a circular  orbit. For both  cases there exist
analytical solutions  to which we  can compare our  numerical results.
In the first case, we have  followed the time evolution of an isolated
$1\,M_{\sun}$  star using  $2\times 10^4$  SPH particles  of  the same
mass.  For the second test we  have followed the evolution of a binary
system made of two white dwarfs  of the same mass ($1\,M_{\sun}$) in a
circular  orbit. Each  one of  the  white dwarfs  was simulated  using
$2\times 10^4$ SPH particles of the same mass.

The  first of our  tests was  designed to  set the  zero point  of our
calculations.  Since  in SPH simulations the particles  are allowed to
move freely under the action  of their own gravitational potential and
of  the  pressure forces,  and  since the  mass  of  each particle  is
relatively large  it is not  obvious {\sl a  priori} whether or  not a
stable    relaxed   configuration   radiates    gravitational   waves.
Figure~\ref{graf_zero} shows the dimensionless strains for the case of
an isolated  white dwarf.   We only show  times larger than  60~s, for
which the  star is  already relaxed to  its final  configuration.  The
relaxation procedure  consisted in allowing  the initial configuration
(which consisted in randomly  distributing the SPH particles according
to the density  profile of a zero temperature white  dwarf of the same
mass) to evolve  for a long enough time until  the oscillations of the
resulting configuration  were completely  negligible.  In this  way we
check  whether  or  not  the  numerical noise  produces  a  negligible
emission of gravitational  waves. And this is indeed  the case.  As it
can be seen  the emission of gravitational waves  is negligible, as it
should  be, given  that the  relaxed configuration  presents spherical
symmetry --- see Eq.~(\ref{mom_quadrupolar}).

\begin{figure}
\centering
\includegraphics[clip,width=250pt]{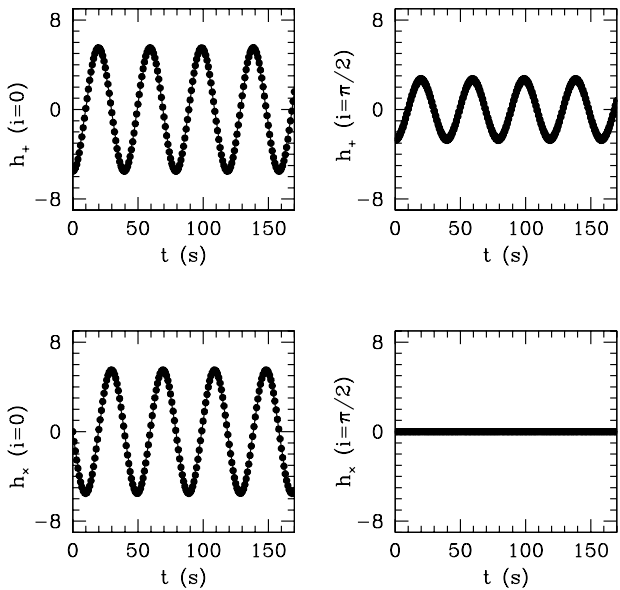}
\caption{Gravitational wave  emission from a close  white dwarf binary
        system.   The dimensionless  strains $h_{+}$  and $h_{\times}$
        are  measured   in  units  of  $10^{-22}$.    The  solid  line
        corresponds  to  the  theoretical  solution whereas  the  dots
        correspond  to our  numerical solution.  Again, the  source is
        assumed to be at a distance of 10~kpc.}
\label{graf_orbit}
\end{figure}

With  regard  to the  second  test, it  is  worth  recalling that  the
emission of gravitational waves from  a binary system, can be obtained
quite  easily  by  assuming   that  both  stars  are  point-like  mass
distributions.   By  doing   so,  from  Eqs.~(\ref{sol_retarded})  and
(\ref{mom_quadrupolar}) one obtains

\begin{equation} 
h_{+} = \sqrt{2} \frac{\mu}{d} \frac{G^{\frac{5}{3}}}{c^4}
(\omega M_{\rm tot})^{\frac{2}{3}}(1+\cos^2 i)\cos  2\omega t
\label{ATT1orbit}
\end{equation}

\noindent and

\begin{equation} 
h_{\times} = 2\sqrt{2} \frac{\mu}{d} \frac{G^{\frac{5}{3}}}{c^4}
(\omega M_{\rm tot})^{\frac{2}{3}}(\noindent \cos i)\sin 2\omega t,
\label{ATT2orbit}
\end{equation}

\noindent where $\omega$  is the angular velocity of  the stars, $\mu$
is the reduced  mass, $M$ is the mass of the white dwarfs,  and $i$ is
the observation angle with respect to the orbital plane.

\begin{figure}
\centering
\includegraphics[clip,width=250pt]{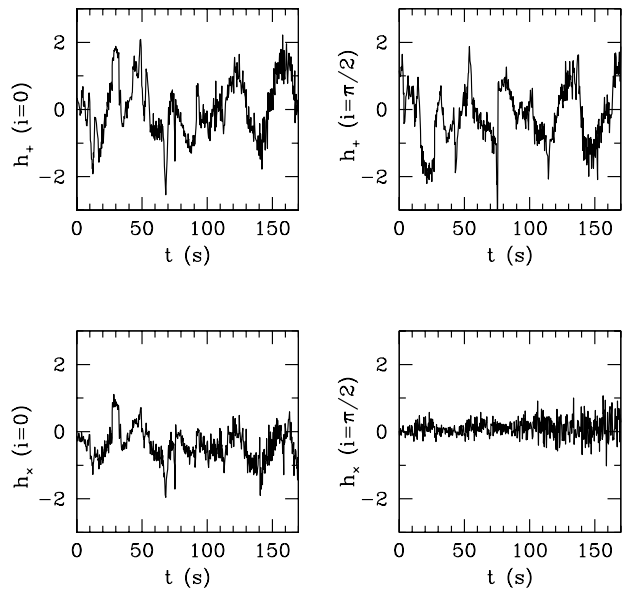}
\caption{Residuals   between   the   theoretical  solution   and   the
        gravitational wave emission from  the close white dwarf binary
        system of Fig.~\ref{graf_orbit}.   Note that the dimensionless
        strains  $h_{+}$ and  $h_{\times}$  are measured  in units  of
        $10^{-23}$, one  order of magnitude smaller than  the scale of
        Fig.~\ref{graf_orbit}.}
\label{difs}
\end{figure}

In our  SPH simulations  we have  chosen the binary  system to  have a
separation  of   $0.05\,  R_{\sun}$  or,   equivalently,  $\omega=7.94
\times10^{-2}\, \mbox{s}^{-1}$. The two white dwarfs describe circular
orbits and no mass  is transferred between both components.  Moreover,
the  two  white  dwarfs  preserve their  initial  spherical  symmetry.
Hence, and according  to Eqs.~(\ref{ATT1orbit}) and (\ref{ATT2orbit}),
we should expect to  obtain dimensionless strains which are sinusoidal
functions with  a period equal to  half of the orbital  period.  As it
can  be  seen  from  Figure~\ref{graf_orbit}, the  numerical  solution
matches  very  well the  theoretical  one.   In  particular, both  the
amplitude and the  frequency ($\nu\simeq\,0.025$~Hz) show an excellent
agreement between theory and  simulations.  To further illustrate this
overall  excellent   agreement,  in  Figure~\ref{difs}   we  show  the
residuals between the theoretical  solution and the gravitational wave
emission  of   the  close  white  dwarf  binary   system  in  circular
orbit. Note  that the  scale in  this case is  one order  of magnitude
smaller than that of  Figure~\ref{graf_orbit}. Hence, we conclude that
our   SPH  simulations   can  accurately   compute  the   emission  of
gravitational waves from coalescing white dwarfs.

\begin{table}
\centering
\caption{Summary  of  the simulations  discussed  in  this paper.  The
number of  SPH particles for the  simulations of the  first section of
the  table is  $4\times  10^4$,  whereas for  the  last simulation  is
$4\times 10^5$.}
\begin{tabular}{lcccc}
\hline
\hline
Run  & $M_{\rm tot}$ ($M_{\sun}$) & Composition & $R_0$ ($R_{\sun}$) & $t$ (s) \\
\hline
1 & 0.4+1.2 & He/ONe & 0.040 &  180 \\
2 & 0.4+0.4 & He/He  & 0.042 &  600 \\
3 & 0.6+0.6 & CO/CO  & 0.041 &  180 \\
4 & 0.6+1.0 & CO/CO  & 0.038 &  725 \\
5 & 0.6+0.8 & CO/CO  & 0.033 &  600 \\
6 & 0.8+1.0 & CO/CO  & 0.028 & 1000 \\
\hline
7 & 0.6+0.6 & CO/CO  & 0.040 &   90 \\
\hline
\hline
\end{tabular}
\end{table}

\section{Gravitational wave emission}

\begin{figure*}
\vspace{16cm}
\includegraphics{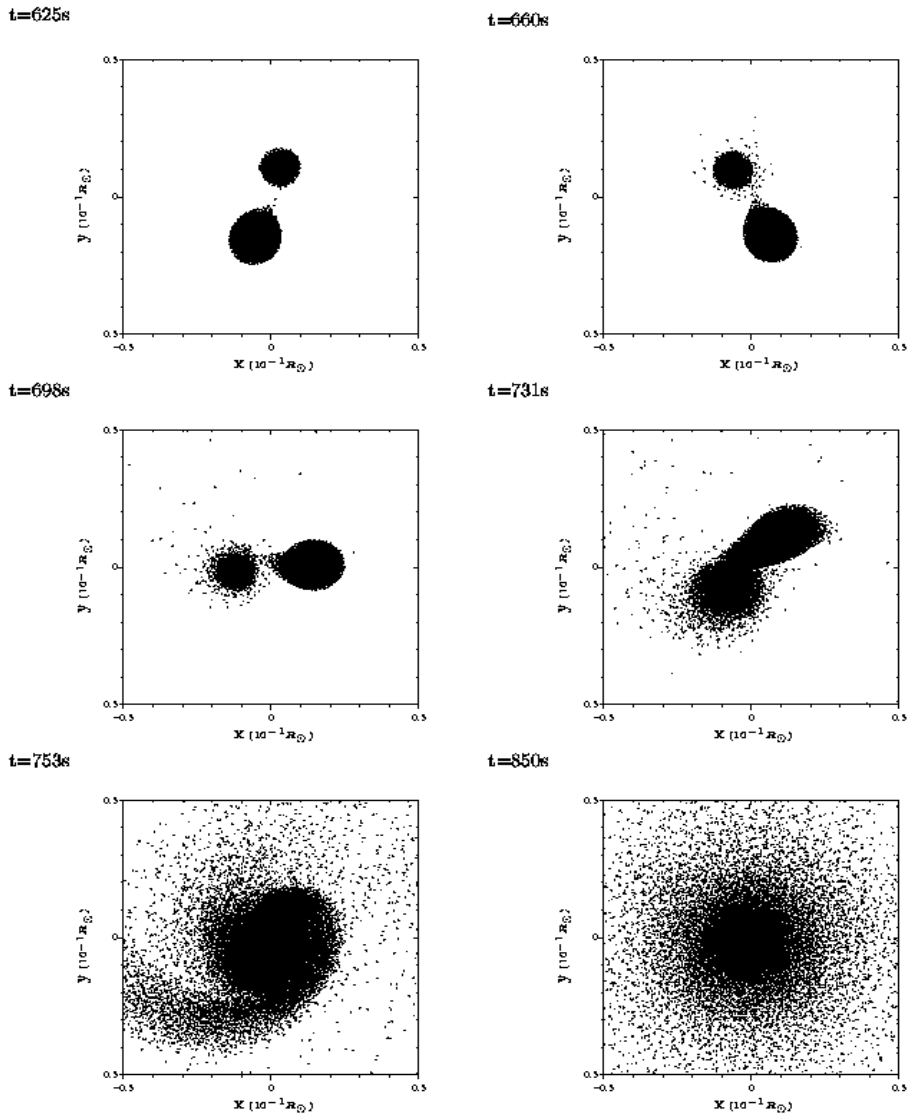}
\caption{Temporal  evolution  of  the 0.8+1.0~$M_{\sun}$  CO-CO  close
        white dwarf binary system  during the most important phases of
        the  merger.  The  SPH particles  have been  projected  in the
        orbital plane. See text for additional details.}
\label{binary}
\end{figure*}

\begin{figure*}
\vspace{16cm}
\includegraphics{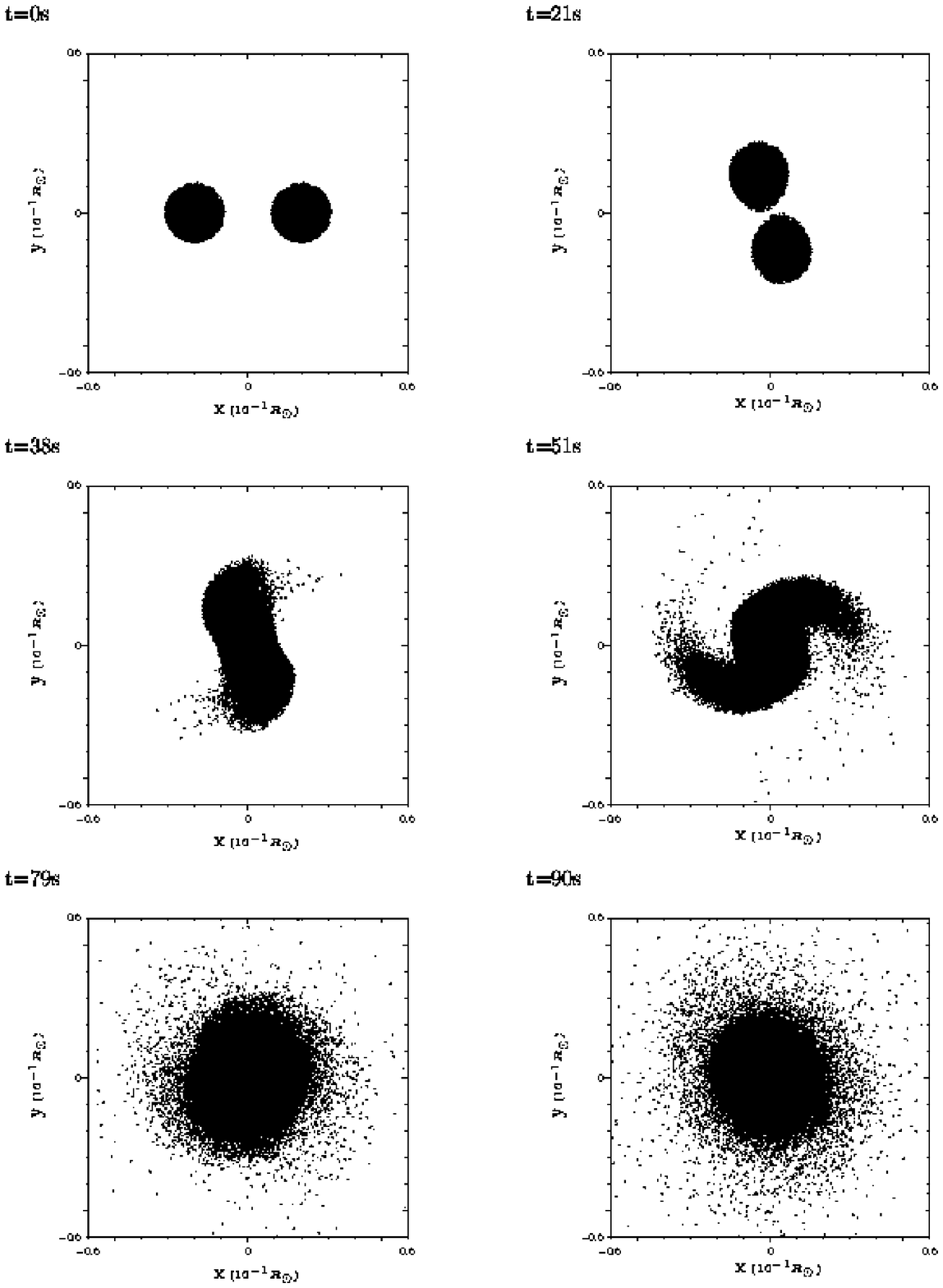}
\caption{Same as figure  \ref{binary} for the 0.6+0.6~$M_{\sun}$ CO-CO
        close white  dwarf binary system  in which $2\times  10^5$ SPH
        particles were used. Only 1 out of 4 particles are shown.}
\label{more}
\end{figure*}

We have  computed the emission  of gravitational waves  resulting from
the  merging  of  several   close  white  dwarf  binary  systems.   In
particular,   the   radiation    of   gravitational   waves   from   a
0.4+0.4~$M_{\sun}$  He-He, a 0.4+1.2~$M_{\sun}$  He-ONe, a  0.6+0.8, a
0.6+1.0  and  a  0.8+1.0~$M_{\sun}$   CO-CO  was  computed.   All  the
simulations performed so far are listed  in Table 1, where the mass of
both components  of the  binary system, their  respective composition,
the  initial  separation  ($R_0$)  and  the  simulation  time  can  be
found. For the sake of conciseness we will only discuss in some detail
the  results  of  the  0.4+0.4~$M_{\sun}$  He-He  system  and  of  the
0.8+1.0~$M_{\sun}$  CO-CO  merger.   In  all  the  cases  the  initial
separation was larger than the  corresponding Roche lobe radius of the
less  massive  component.  For  instance,  for  the  case in  which  a
0.4+0.4~$M_{\sun}$ He-He  merger is considered  the initial separation
was $\simeq 0.042\,R_{\sun}$, and  in the case of a 0.8+1.0~$M_{\sun}$
CO-CO  system  the initial  separation  was $\simeq  0.027\,R_{\sun}$.
Instead  of computing  self-consistently  the chirping  phase we  have
chosen to add  a very small artificial radial  acceleration term which
decreases  the separation  of both  components until  the  last stable
orbit  is  reached. This  acceleration  term  is  proportional to  the
velocity,  never  amounts to  more  than a  5\%  of  the real  orbital
acceleration  and is added  once the  stars have  already done  a full
orbit, then we let the system relax during another orbit and the whole
procedure is repeated again until  the secondary fills its Roche lobe.
Once  the secondary  fills its  Roche lobe  this acceleration  term is
suppressed and  the system is allowed to  evolve freely. Nevertheless,
we have  checked (see  \S 6) that  the amplitude of  the gravitational
waves during the initial phase  of the coalescence agrees with that of
the chirping phase.  Finally, it  is important to mention here that in
all  the  simulations  studied  in  the present  work  the  number  of
particles   for  each   white   dwarf  is   $2\times   10^4$  in   all
cases. However, and  in order to check the  sensitivity of our results
to the  number of particles, we  have run an  additional simulation in
which the number of  particles was significantly increased to $2\times
10^5$ for  each star, this simulation  is listed in the  last entry of
table 1 and discussed below.

\begin{figure}
\centering
\includegraphics[clip,width=250pt]{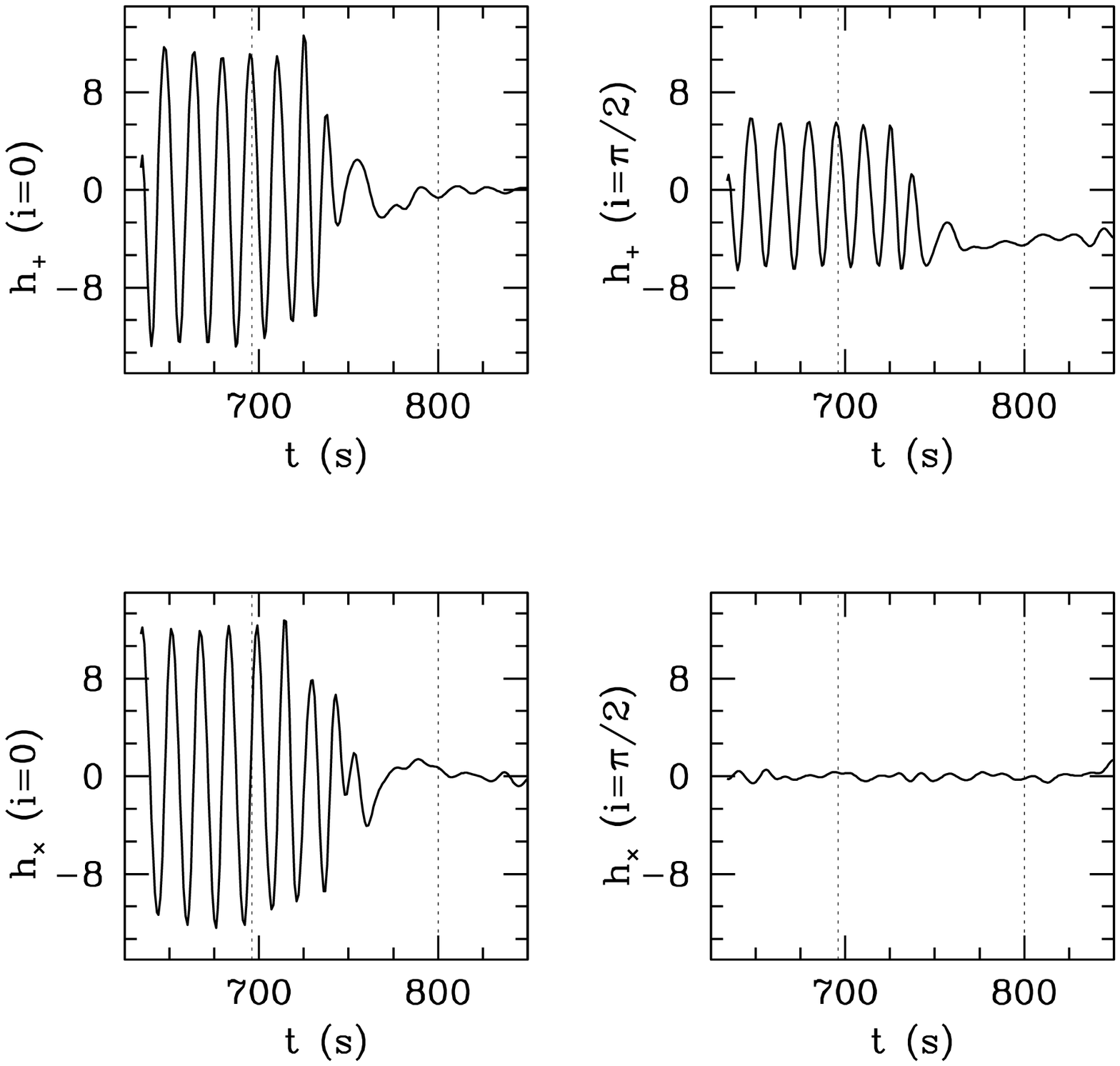}
\caption{Gravitational   wave   emission   from   the  merger   of   a
        0.8+1.0~$M_{\sun}$   CO-CO close  white  dwarf binary  system.
        The  dimensionless   strains  $h_{+}$  and   $h_{\times}$  are
        measured  in units  of $10^{-22}$.   The leftmost  thin dotted
        line corresponds  to the  time at which  the two  white dwarfs
        start to merge and the rightmost thin  dotted line corresponds
        to the time at which an approximate  cylindrical configuration
        has been already achieved.  Again, the source is assumed to be
        at a distance of 10~kpc.}
\label{graf_hmerger1}
\end{figure}

In Fig.~\ref{binary}  we show the temporal evolution  of the positions
of the SPH  particles projected on to the orbital  plane as a function
of  time  for  the  $0.8   +  1.0\,  M_{\sun}$  system.   The  initial
configuration of  the two stars was  completely spherically symmetric.
After some time the secondary is tidally deformed (top left panel) and
begins to overflow its Roche lobe (top right panel). As a consequence,
an accretion stream forms  (middle left panel).  This accretion stream
is  directed  towards the  primary,  forming  an  arm.  The  particles
flowing from the secondary onto the primary are redistributed over the
surface of the primary (middle right panel) and the arm twists as time
increases (bottom  left panel),  leading to the  formation of  a heavy
accretion disk  with cylindrical  symmetry (bottom right  panel).  The
whole process lasts for about 3.8 minutes.  The merging process can be
understood  in  terms of  the  positive  feedback  experienced by  the
secondary: as the coalescence  proceeds the secondary looses mass and,
thus, becomes less dense and  expands leading to an enhanced mass-loss
rate which, in turn, leads to a decrease of the average density of the
secondary.   Very few  particles  achieve velocities  larger than  the
escape velocity and, hence, a very small fraction of the total mass is
ejected from the system (about $6.5\times 10^{-3} \, M_{\sun}$ in this
case).  An important  feature of the simulation is  that the accretion
disk  is supported  by its  own rotational  velocity.  However,  it is
important to  realize that  in the final  configuration a  weak spiral
pattern can still be found.   This pattern should become less and less
apparent as  time increases, reaching cylindrical  symmetry during the
very late stages of the  simulation.  However, following the long term
evolution  of  this  heavy   accretion  disk  would  require  a  heavy
computational  load which  is well  beyond our  current possibilities.
Additionally, the  accretion rate onto the  primary becomes negligible
during  this  phase. Hence,  our  final  configuration  consists in  a
central rotating  spherically symmetric  compact star surrounded  by a
keplerian  and  almost  cylindrically  symmetric  accretion  disk.  As
discussed in  Guerrero et  al. (2004) the  rotational velocity  of the
central star has  been problably overestimated due to  the large shear
introduced by  SPH methods. We  have used the artificial  viscosity of
Balsara  (1995),  which  does  not  produce an  excessive  shear  and,
consequently,  reduces somewhat (but  not completely)  these problems.
Therefore, some properties  of the merged object could  be affected by
the excess  of rotation of the  primary.  However, since  even in this
case  the  star preserves  spherical  symmetry  we  consider that  the
calculations  described  below provide  a  good  approximation to  the
emission of  gravitational waves.  All the cases  studied present more
or less  the same features except  those in which two  white dwarfs of
equal mass are  involved. In such a case the  final configuration is a
single  spheroidal central  object ---  see, for  instance,  Fig.~5 of
Guerrero et al. (2004).

In   figure~\ref{more}  the   coalescence  of   a  binary   system  of
0.6+0.6~$M_{\sun}$  in which  each  of the  stars  was modelled  using
$2\times 10^5$  particles is  displayed. As in  figure~\ref{binary} we
have chosen to  represent the temporal evolution of  the SPH particles
projected in the orbital plane.  Note, however, that in this case only
one out of four particles has been represented. In this case, however,
the simulation  only covers times larger  than that at  which the last
stable orbit of the system occurs.  As it can be seen, the results are
essentially the same  and, thus, we are confident  in the main results
and  general trends of  our numerical  simulations. We  will, however,
come  back later  to  this issue  at  the end  of  this section,  when
discussing the emission of gravitational waves.

\begin{figure}
\centering
\includegraphics[clip,width=250pt]{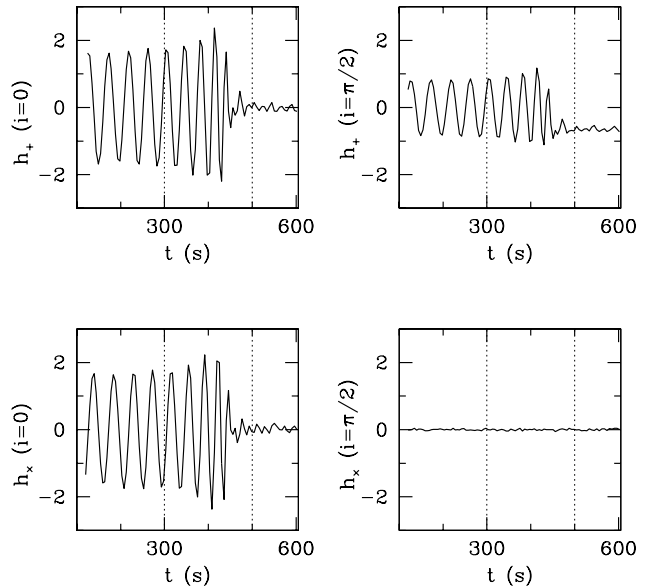}
\caption{Same  as  Fig.~\ref{graf_hmerger1}  but  for the  case  of  a
        0.4+0.4~$M_{\sun}$ He-He close white dwarf binary system.}
\label{graf_hmerger2}
\end{figure}

An example of our  results is shown in figures~\ref{graf_hmerger1} and
\ref{graf_hmerger2},  where   the  dimensionless  strains   $h_+$  and
$h_\times$  as  a function  of  time  for  different inclinations  are
respectively   shown  for   the  0.8+1.0~$M_{\sun}$   CO-CO   and  the
0.4+0.4~$M_{\sun}$ He-He systems.  The beginning and the final time of
the merging itself are shown in  both figures as thin dotted lines. In
figure~\ref{graf_hmerger1} it can be  seen that before the coalescence
proceeds the  emission of gravitational  waves still has  a sinusoidal
pattern, but with  an increasing frequency.  That is,  the close white
dwarf binary system  chirps as a consequence of  the spiral trajectory
of the stars  towards the center of mass.   Note that $h_{\times}$ for
$i=\pi/2$ is zero because the orbital plane is parallel to the line of
sight.  When the two white  dwarfs start to coalesce, the amplitude of
the dimensionless strains somehow  increase first, but only during the
first (and most violent) stage of the merger.  This corresponds to the
phase in  which a spiral  arm is formed.  After the second  maximum is
achieved the amplitude decreases dramatically.  In fact, only two more
clear maxima can be apparently distinguished before the system reaches
its  final configuration.   It is  interesting to  note that  once the
merger has already finished one  of the dimensionless strains still is
significant,  $h_+  \simeq  -5   \cdot  10^{-22}$  at  $d=10$~kpc  for
$i=\pi/2$.  This  residual emission is  due to the  inhomegenieties of
the accretion  disk previously discussed.  Each one of the  very small
maxima  appearing  at  very   late  times  corresponds  to  successive
crossings of the edge  of the weak spiral arm in front  of the line of
sight. Note  as well  that this residual  emission tends  to disappear
assymptotically, as a consequence  of the ongoing rehomogeneization of
the  accretion   disk.   It  is   important  to  realize   that  these
inhomogeneities could be either an artifact due to the resolution used
in  our SPH  simulations or  a consequence  of the  adopted artificial
viscosity, since  it is  well known that  the artificial  viscosity of
Balsara (1995) induces a considerable shear viscosity.

In figure \ref{graf_hmerger2} it can  be seen that although during the
first part of simulation the  same chirping pattern is found, once the
merger proceeds the gravitational wave signal suddenly disappears on a
short time scale, comparable to the orbital period. This is due to the
fact that in this case the two stars have equal masses and, therefore,
there  is not  a prominent  spiral accretion  stream. Instead  the two
components of  the binary  system are disrupted  around the  center of
masses, where  a shocked region forms  as a consequence  of the impact
between two streams.   The typical Mach numbers in  the shocked region
are Ma~$\sim 1$. The streams are slightly asymmetric, depending on the
orbital phase  at which they form.   By the end of  the simulation ---
see again  Fig.~5 of Guerrero et  al.  (2004) ---  both streams become
entangled and the final configuration of the resulting object consists
in  a central  shocked  region  surrounded by  a  less dense  rotating
spheroid, in  which a  certain degree of  asymmetry is  still present.
Hence,  the coalescing  process does  not present  a  rather symmetric
behavior  during the initial  phases and,  consequently, we  see three
consecutive maxima.  After this, once the streams become entangled the
emission of gravitational waves is  heavily suppressed and, by the end
of  the  merging  process,  the  emission of  gravitational  waves  is
negligible.  However, note as well  that in this case a small residual
emission is  also observed in the  dimensionless strain $h_+=-0.8\cdot
10^{-22}$ at  $d=10$~kpc for $i=\pi/2$,  although considerably smaller
than  in  the  previous  case.   This  is again  due  to  our  limited
computational resources.  The impossibility of following the very late
phases of  the coalescence  episode does not  allow us to  compute the
rehomogeneization of the external  spheroid and, hence, to compute the
long-term behavior of $h_+$ accurately.

\begin{figure}
\centering
\includegraphics[clip,width=250pt]{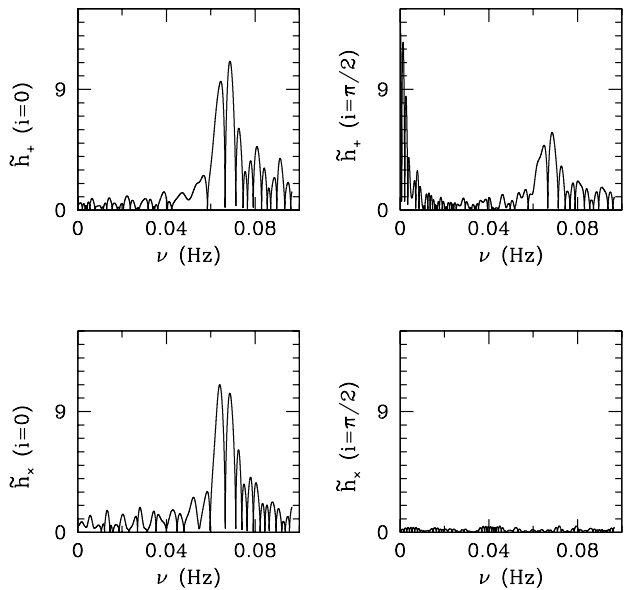}
\caption{Modulus  of the  Fast Fourier  Transform of  the adimensional
        strains $h_{+}$ and $h_{\times}$ of Fig.~\ref{graf_hmerger1}.}
\label{graf_fftmerger1}
\end{figure}

\begin{figure}
\centering
\includegraphics[clip,width=250pt]{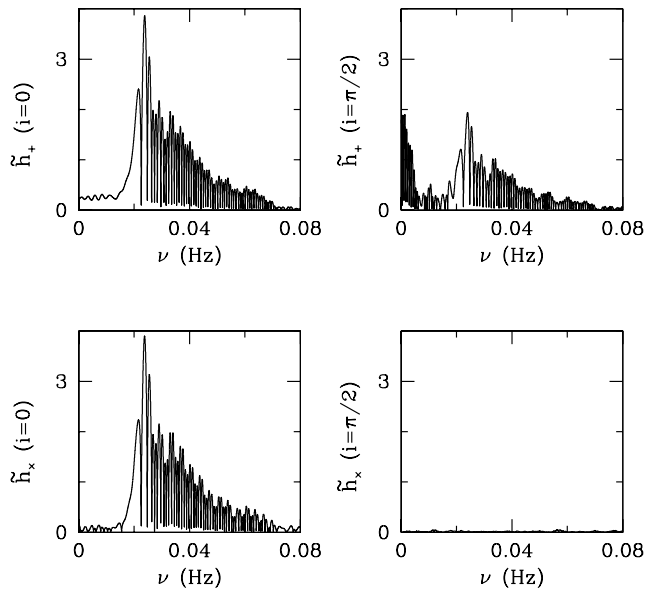}
\caption{Modulus  of the  Fast Fourier  Transform of  the adimensional
        strains $h_{+}$ and $h_{\times}$ of Fig.~\ref{graf_hmerger2}.}
\label{graf_fftmerger2}
\end{figure}

The same information is displayed in figures~\ref{graf_fftmerger1} and
\ref{graf_fftmerger2},  but in  a different  format.  As  can  be seen
there,  the  dominant  frequency  is  given  by  the  orbital  period.
However, as the  components of the binary system  approach each other,
the dominant  frequency is  shifted to larger  values and,  during the
merger, high frequencies show up, although with very low amplitudes.

Finally, and in  order to check the sensitivity of  our results to the
resolution of the SPH simulations described above we have computed the
emission  of gravitational  waves for  both  our run  ``7'', in  which
$2\times 10^5$ particles  were used for each star,  and our run number
3, in which a resolution 10  times poorer was adopted. The results are
displayed in Fig.~\ref{lr-hr}. As in the previous figures the vertical
thin  lines  denote the  moment  at which  the  last  stable orbit  is
achieved.  Note, nevertheless, that the high resolution simulation was
started, as previously  mentioned, when the system was  already at the
last stable orbit. That is the  reason why the chirping phase does not
appear.  Consequently, the  time origin  of this  simulation  has been
shifted accordingly to match that of the low resolution simulation. As
it can be seen the emission  of gravitational waves is very similar in
both cases, and, hence, our results are robust.

\section{Detectability}

\begin{table*}
\caption{Maximum distance  at which LISA  will detect the  close white
dwarf binary systems discussed in this paper, see text for details.}
\centering
\begin{tabular}{lccccccc}
\hline
\hline
Run & $M_{\rm tot}$ &$\nu_0$ &$M$ &$d_{\rm max}$ & $\eta$ & $h_{\rm max}$  & $E$\\
& $(M_{\sun})$ & (mHz)  & ($M_{\sun}$) & (kpc)        & (10 kpc) & ($10^{-22}$) & ($10^{41}$~erg)\\
\hline
1 & 0.4+1.2 & 32 & 0.59 & 21 & 10.5 & 11.4 &   6\\
2 & 0.4+0.4 & 21 & 0.35 & 10 & 5.0  &  2.4 & 0.1\\
3 & 0.6+0.6 & 13 & 0.53 & 26 & 13.0 &  6.0 & 0.6\\
4 & 0.6+1.0 & 34 & 0.67 & 29 & 14.5 &  6.8 &   2\\
5 & 0.6+0.8 & 40 & 0.60 & 31 & 15.6 &  6.1 &   1\\  
6 & 0.8+1.0 & 58 & 0.77 & 33 & 16.6 & 12.8 &   7\\
\hline
7 & 0.6+0.6 & 13 & 0.53 & 26 & 12.8 &  5.9 & 0.6\\
\hline
\hline
\end{tabular}
\end{table*}

In order to check whether or not  LISA would be able to detect a close
white  dwarf binary  system  we  have proceeded  as  follows. We  have
already shown that the most prominent feature of the emitted signal is
its sudden disappearance  in a couple of orbital  periods and that the
gravitational  wave emission  during  the coalescence  phase does  not
increase  noticeably.   Hence,  the  gravitational  wave  emission  is
dominated  by the  chirping phase.   Hence, we  have assumed  that the
orbital  separation of the  two white  dwarfs is  exactly that  of our
binary  system {\sl  when  mass  transfer starts}.   We  have done  so
because,  as  explained  before,  we  have added  a  small  artificial
acceleration term  to the initial  configuration in order to  avoid an
excessive  computational   demand  at   the  very  beginning   of  our
simulations.  This acceleration term  is suppressed once the secondary
begins to transfer mass onto the primary. Note, however, that the mass
transfer  starts  when  the   secondary  fills  its  Roche  lobe  and,
consequently, this  orbital separation  is physically sound.   We have
further assumed  that the integration time  of LISA will  be one year.
We have checked  that during this period the  variation of the orbital
separation  is negligible  (see  also figures~\ref{graf_hmerger1}  and
\ref{graf_hmerger2}).   Of  course,  should  the integration  time  be
smaller the signal-to-noise ratio  derived below would smaller. Hence,
our results should be regarded as an upper limit.

The signal-to-noise ratio, $\eta$, is given by

\begin{equation}
\eta^2=\int_{-\infty}^{+\infty}\frac{\tilde{h}^2(\omega)}{S(\omega)}
\frac{d\omega}{2\pi}
\end{equation}

\noindent where  $S(\omega)=S_{\rm h}(\omega)\tau$ is  the sensibility
of LISA, $\tau$ is  the integration period, and $\tilde{h}(\omega)$ is
the Fourier Transform  of the dimensionless strain.  It  can be easily
shown     that    for     a     monocromatic    gravitational     wave
$\eta=h(\omega)/{S_{\rm  h}^{1/2}(\omega)}$.   The  maximum  distance,
$d_{\rm max}$,  at which LISA  would be able  to detect a  close white
dwarf binary system is then:

\begin{figure}
\centering
\includegraphics[clip,width=250pt]{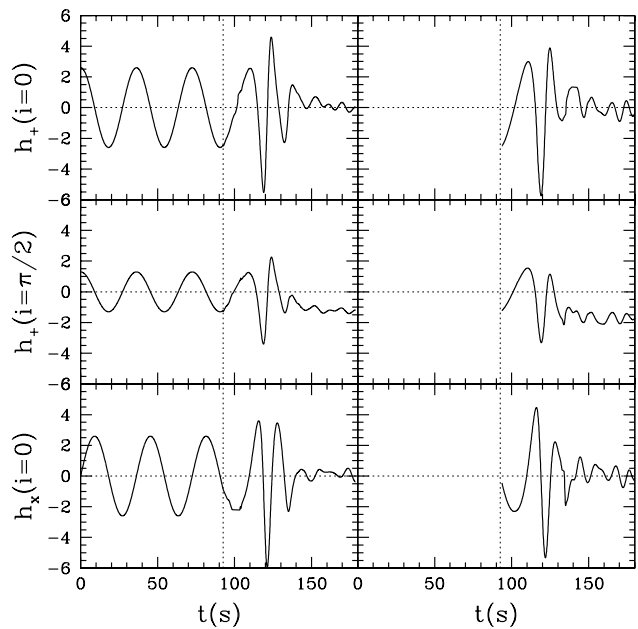}
\caption{A comparison of the  computed emission of gravitational waves
         when the  resolution of our  SPH simulations is  changed. The
         left panels  show the emission  of gravitational waves  for a
         simulation  in which  a  low resolution  ($2\times 10^4$  SPH
         particles)  was  used.   The   right  panels  show  the  same
         quantities when a high resolution was adopted ($2\times 10^5$
         SPH particles). See text for further details.}
\label{lr-hr}
\end{figure}

\begin{equation} 
d_{\rm max} \approx 17 \big(\frac{5}{\eta}\big)  
\big(\frac{M}{M_ {\sun}}\big)^{5/3} 
\big(\frac{\nu_0}{1\; {\rm mHz}}\big)^{2/3} 
\big(\frac{10^{-23}}{\sqrt{S_{\rm h}(\nu_ 0)}}\big)  
\; {\rm kpc}
\end{equation}

\noindent where  $\nu_0$ is the  frequency of gravitational  wave, and
$M$ is the chirping mass.

\begin{equation}
M=(\mu M_{\rm tot}^{2/3})^{3/5}
\end{equation}

\noindent  being  $\mu=m_1  m_2/M_{\rm  tot}$ the  reduced  mass,  and
$M_{\rm tot}=m_1+m_2$ the total mass. 

\begin{figure}
\centering
\includegraphics[clip,width=250pt]{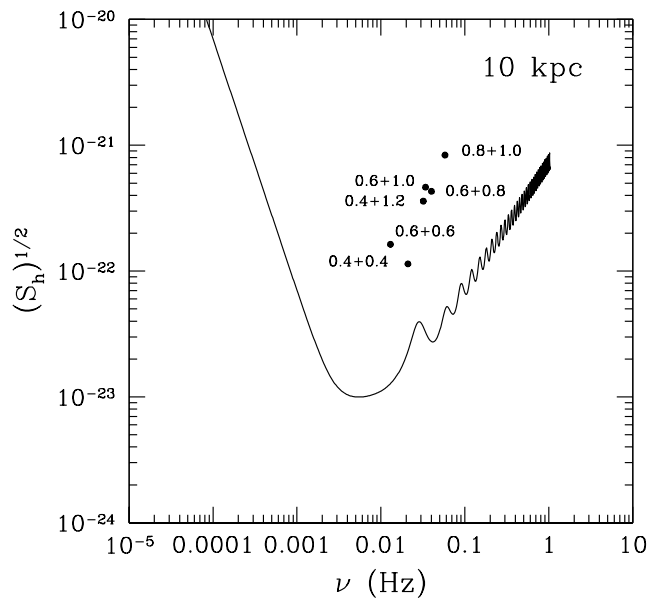}
\caption{A comparison of the signal  produced by the close white dwarf
        binary  systems studied  here, when  a distance  of  10~kpc is
        adopted, with the spectral distribution of noise of LISA for a
        one year integration period.}
\label{graf_snr}
\end{figure}

In order to evaluate the maximum  distance at which LISA would be able
to detect  the close white dwarf  binary systems studied  here we have
adopted $\eta=5$.  We have furthermore used the integrated sensibility
of         LISA          as         obtained         from         {\tt
http://www.srl.caltech.edu/$\sim$shane/sensitivity}.   The results are
given in Table 2, where the  frequency of the close white dwarf binary
system when the secondary overflows its Roche lobe, the chirping mass,
the  maximum  distance   at  which  LISA  would  detect   them  for  a
signal-to-noise  ratio  of  5   and  one  year  integration,  and  the
signal-to-noise ratio  at 10~kpc for  one year integration  are shown.
Also  shown  in Table  2  are  the peak  amplitude  at  10~kpc of  the
dimensionless strain  (in units of $10^{-22}$) and  the total radiated
energy  in the form  of gravitational  waves.  We  stress that  if the
merger   occurs   during  the   one   year   integration  period   the
signal-to-noise ratio would  be smaller. It is worth  noticing as well
that  the energy  radiated  away  from the  binary  system during  the
coalescence  in the form  of gravitational  waves is  of the  order of
$\sim 10^{41}$~erg, much  smaller than the total energy  of the system
and,  hence, totally  negligible  in  the energy  budget,  so no  back
reaction should be  taken into account and the  procedure used here is
robust. Note as well, that for the case of $0.6+0.6\,M_{\sun}$ system,
in which two resolutions were  used the maximum distance at which LISA
would be able  to detect the coalescence is very  similar for both the
high   resolution  simulation,   12.8~pc,  and   the   low  resolution
simulation, 13.0~pc.   The energy  released during the  coalescence is
very  similar   as  well  in   both  cases,  0.63  and   $0.64  \times
10^{41}$~erg, respectively.

Finally, in Fig.~\ref{graf_snr} we  compare the signal produced by the
close  white  dwarf binary  systems  studied  in  this paper,  when  a
distance of 10~kpc is adopted, with the spectral distribution of noise
of LISA for a  one year integration period. As it can  be seen, all of
them will be eventually detected, at different signal-to-noise ratios,
ranging from  $\sim 5.0$ for  the $0.4+0.4\,M_{\sun}$ system  to $\sim
16.6$ for the $0.8+1.0\,M_{\sun}$ system.

\section{Discussion and conclusions}

We have computed the emission  of gravitational waves of merging white
dwarf binaries,  for a  wide range of  masses and compositions  of the
components of the  binary system. For that purpose we  have used a SPH
code  which  allowed  us  to  follow the  temporal  evolution  of  the
coalescing  white dwarfs.   We  have shown  that  the most  noticeable
feature  of  the emitted  signal  is  a  sudden disappearance  of  the
gravitational strains.  By contrast  the chirping phase will be easily
detectable by  future space-borne interferometers like  LISA. In fact,
it can be said that the  most relevant signature of the merger will be
the  absence of  any signature  and  the sudden  disappearance of  the
source. Since the frequency coverage of LISA will range from $10^{-1}$
to $10^{-4}$~Hz,  the detection of  chirping close white  dwarf binary
systems  is  guaranteed  (Farmer  \&  Phinney 2003;  Nelemans  et  al.
2001a).   Moreover, at  frequencies of  $\sim 3\cdot  10^{-3}$~Hz most
close white dwarf binary  systems will be spectrally resolved (Cornish
\& Larson  2003). Typically, LISA will  be able to  detect $\sim 3000$
binaries at  these frequencies (Seto 2002).  Additonally,  it has been
recently shown  (Cooray, Farmer \&  Seto 2004) that the  typical error
box  of LISA  for  these  kind of  systems  will be  $\delta\Omega\sim
5$~deg$^2$, and that, given that  many of the detected sources will be
eclipsing binaries  with a period  equal to that of  the gravitational
waves, optical follow-up campaigns  will allow us to further constrain
the location of the sources.  Hence, by combining optical observations
and  gravitational  wave  data,  and  taking  into  account  that  the
gravitational wave  signal for such mergers suddenly  vanishes --- or,
equivalently, that  during the integration  period the signal-to-noise
ratio stops  growing --- we  should be able  to gain insight  into the
physics of merging and to obtain precise information of the properties
of the progenitor  systems. Moreover, it should be  taken into account
that for a typical merger ---  namely, the $0.6 + 0.8\, M_{\sun}$ case
--- the  volume  accesible  to  LISA is  $V_{\rm  LISA}\sim  1.2\times
10^{14}$~pc$^3$.   Since the volume  of the  Galaxy is  $V\sim 3\times
10^{11}$~pc$^3$, LISA would be able to detect all the mergers occuring
in our Galaxy during its  operation period.  However, the typical rate
of  white  dwarf  mergers   is  $r\sim  8.3  \times 10^{-3}$~yr$^{-1}$
(Nelemans  2003) and,  hence, although  there is  an uncertainty  of a
factor of 5 in the rate of white dwarf mergers, the expected detection
rate is consequently small.

\vspace{0.5 cm}

\noindent  {\sl  Acknowledgements.}   This  work  has  been  partially
supported   by   the   MCYT  grants   AYA2002--4094--C03--01/02,   and
AE-ESP2002-10451-E,  by the  European  Union FEDER  funds  and by  the
CIRIT.  We  would also  like to acknowledge  the invaluable  advise of
J.M.   Ib\'a\~nez   who  largely  contributed   through  his  support,
suggestions and comments to improve  the manuscript. We also thank our
anonymous referee for valuable comments and criticisms.

\end{document}